\documentclass[twocolumn, aps, prx, reprint, superscriptaddress]{revtex4-2}
\usepackage{newtxtext}
\usepackage{amssymb}  
\usepackage{amsfonts}  
\usepackage{amsmath}  

\usepackage{amsthm}  
\usepackage{bm}  

\usepackage{bbm}  
\usepackage{xcolor}
\usepackage[colorlinks, allcolors=blue]{hyperref}
\usepackage{graphicx}
\usepackage{tabularx}
\usepackage{bbding}
\usepackage{comment}
\usepackage{float}
\usepackage{natbib}
\usepackage{ifthen}
\usepackage{cleveref}
\usepackage{lipsum}
\usepackage{makecell}
\usepackage{braket}
\usepackage[english]{babel}
\usepackage[autostyle, english = american]{csquotes}
\MakeOuterQuote{"}

\theoremstyle{definition}

\theoremstyle{remark}

\usepackage{tikz, datatool, xcolor, ifthen, csvsimple}
\usepackage[framemethod=TikZ]{mdframed}
\usetikzlibrary{arrows, calc, shapes, math, decorations, decorations.markings}
\usetikzlibrary{positioning}
\usetikzlibrary{shapes.geometric}
\usetikzlibrary {arrows.meta}
\usepackage{pgfplots}
\pgfplotsset{compat=newest}
\pgfplotsset{
    colormap={mycm}{rgb255=(225, 225, 225) rgb255=(225, 225, 225)},
    colormap/mycm/.style={
        colormap name=mycm,
    },
}
\usepackage{pgfmath}
\input pgfmath.tex
\tikzset{midarrow/.style={
    decoration={markings, mark=at position 0.55 with {\arrow{>}}},
    postaction={decorate}
}}
\tikzset{midarrowrev/.style={
    decoration={markings, mark=at position 0.45 with {\arrow{<}}},
    postaction={decorate}
}}

\definecolor{green}{RGB}{50, 180, 50}
\definecolor{blue}{RGB}{20, 30, 250}

\begin{document}

\title{Revealing quantum phase string effect in doped Mott-insulator: a tensor network state approach}

\author{Wayne Zheng}
\affiliation{Department of Physics, The Chinese University of Hong Kong, Sha Tin, New Territories, Hong Kong, China}
\affiliation{Department of Physics, Southern University of Science and Technology, Shenzhen 518055, China}

\author{Jia-Xin Zhang}
\affiliation{French American Center for Theoretical Science, CNRS, KITP, Santa Barbara, California 93106-4030, USA
}
\affiliation{Kavli Institute for Theoretical Physics, University of California, Santa Barbara, California 93106-4030, USA}

\author{Zheng-Yuan Yue}
\affiliation{Department of Physics, The Chinese University of Hong Kong, Sha Tin, New Territories, Hong Kong, China}

\author{Zheng-Cheng Gu}
\affiliation{Department of Physics, The Chinese University of Hong Kong, Sha Tin, New Territories, Hong Kong, China}

\author{Zheng-Yu Weng}
\affiliation{Institute for Advanced Study, Tsinghua University, Beijing 100084, China}

\date{\today}

\begin{abstract}

We apply the fermionic tensor network (TN) state method to understand the strongly correlated nature in a doped Mott insulator. We conduct a comparative study of the $\sigma{t}$-$J$ model, in which the no-double-occupancy constraint remains unchanged but the quantum phase string effect associated with doped holes is precisely switched off.
Thus, the ground state of the $\sigma{t}$-$J$ model can serve as a well-controlled reference state of the standard $t$-$J$ model.
In the absence of phase string, the spin long-range antiferromagnetic (AFM)  order is found to be essentially decoupled from the doped holes, and the latter contribute to a Fermi-liquid-like compressibility and a coherent single-particle propagation with a markedly reduced pairing tendency.
In contrast, our TN calculations of the $t$-$J$ model indicate that the AFM order decreases much faster with doping and the single-particle propagation of doped holes gets substantially suppressed, concurrently with a much stronger charge compressibility at small doping and a significantly amplified Cooper pairing tendencies.
These findings demonstrate that quantum many-body interference from phase strings plays a pivotal role in the $t$-$J$ model, mediating long-range entanglement between spin and charge degrees of freedom.


\end{abstract}

\maketitle

\section{Introduction}

The discovery of the high-temperature (high-$T_{c}$) superconductivity in copper oxides stands as one of the major breakthroughs in condensed matter physics in last century~\cite{Bednorz1986, RevModPhys.66.763}.
Despite the passage of time, it is still under active investigations both experimentally and theoretically.
Shortly after its discovery, it is realized that the high-$T_{c}$ physics mostly settles to doping a Mott insulator, which is widely believed to be described by the effective one-band $t$-$J$ model~\cite{BASKARAN1987973, PhysRevB.37.3759, RevModPhys.78.17}.
Its Hamiltonian reads $H_{t\text{-}J}\equiv{H}_{t}+H_{J}$, in which
\begin{equation}
    \begin{aligned}
    H_{t}&=-t\sum_{\langle{ij}\rangle,\sigma}\left(
        {c}_{i\sigma}^{\dagger}{c}_{j\sigma}+h.c.
    \right), \\
    H_{J}&=J\sum_{\langle{ij}\rangle}\left(
        \mathbf{S}_{i}\cdot\mathbf{S}_{j}-\frac{n_{i}n_{j}}{4}
    \right).
    \end{aligned}
    \label{eq:ham_tJ}
\end{equation}
$n_{j}\equiv\sum_{\sigma}c_{j\sigma}^{\dagger}c_{j\sigma}$ is the particle number operator on site $j$.
$\langle{ij}\rangle$ denotes a nearest-neighbor (NN) bond.
The spin-1/2 operator is given by 
$\mathbf{S}_i=(1/2)\sum_{\alpha,\beta}c^\dagger_{i\alpha} \boldsymbol{\sigma}_{\alpha \beta} c_{i\beta}$, 
where $\boldsymbol{\sigma} = (\sigma^x, \sigma^y, \sigma^z)$ are the Pauli matrices.
A Mott insulator is described by the antiferromagnetic (AFM) Heisenberg model $H_{J}$ where electrons are all localized on site due to strong Coulomb interactions.
Upon hole doping, electrons are removed and doped holes can move as spin-$1/2$ fermions described by $H_{t}$.
These dopants introduce novel physics that has yet to be fully understood.
At present, we still cannot have a complete theoretical understanding on this seemingly simple model.
It is nonintegrable, strongly correlated and plagued by a severe sign-problem~\cite{PhysRevLett.94.170201, RevModPhys.94.015006}.

While Landau's Fermi liquid theory~\cite{Landau1956zuh, baym2008landau} successfully describes certain strongly interacting fermions (e.g. conventional metals),
its foundational premise—that interacting bare particles can be fully renormalized into \emph{quasiparticles} with conserved spin and charge, maintaining a one-to-one correspondence near the Fermi surface—fails dramatically in copper oxides.
This is evidenced by stark contradictions particularly in the temperature-dependent transport properties of the strange metal phase~\cite{Keimer2015}.
Nowadays, it is widely accepted that the Cooper oxide high-$T_{c}$ superconductivity largely originates from a non-Fermi liquid quantum matter~\cite{PhysRevB.81.184515, Samuel2017, hartnoll2018} and described by the $t$-$J$ model.
Historically, the theoretical foundation for understanding this can be traced back to a one-dimensional (1D) Luttinger liquid.
Haldane emphasized that the key distinction between a Luttinger liquid and a Fermi liquid is the presence of \emph{unrenormalizable} forbidden scattering regions in the low-energy excitations of the former~\cite{Haldane1981}.
Later, Anderson proposed a potential Luttinger-like liquid in a two-dimensional (2D) Hubbard model~\cite{PhysRevLett.64.1839}, which stems from a unrenormalizable Fermi-surface phase shift.
Remarkably, this idea was sharply identified and formulated in the $t$-$J$ model, known as the \emph{phase string effect}~\cite{PhysRevLett.77.5102, PhysRevB.55.3894}.
Growing evidence suggests that these special quantum Berry phases underlie both the breakdown of Landau quasiparticles in doped Mott insulators and the central mysteries of cuprate physics~\cite{Zaanen2011, PhysRevB.98.165102, PhysRevX.12.011062}.
These quantum phase strings \emph{cannot} be renormalized into the effective magnetic moment or mass of bare local electrons any longer.
Instead, it can only be written as a dressed non-local object
\begin{equation}
    \Tilde{c}_{j\sigma}\equiv{c}_{j\sigma}e^{-\text{i}\Omega_{j}},
\end{equation}
where $\Omega_{j}$ represents accumulated non-local phases induced by the doped hole~\cite{PhysRevLett.80.5401, PhysRevB.59.8943}.
Moreover, another canonical example of a 2D non-Fermi liquid is the fractional quantum Hall liquid (FQHL) enriched with topological orders~\cite{Wen1995, PhysRevB.61.10267}.
In a FQHL, quasiparticles carry fractional charge and follow a fractional statistics.
Notably, these non-Fermi liquids share a fundamental theoretical similarity:
\emph{their elementary excitations do not trivially correspond to the original bare electrons but instead emerge through specific, non-perturbative mechanisms.}
This breakdown of the one-to-one correspondence highlights the profound departure from conventional paradigms.
The mapping between bare particles and emergent excitations in such systems still relies heavily on case-specific insights~\cite{PhysRevLett.50.1395, Zaanen2011}.
A unified analytical framework for constructing all types of non-Fermi liquids remains elusive.

On the other hand, the tensor network (TN) method, originating from modern quantum entanglement and information theory, has become increasingly important for both numerical simulations and theoretical analyses towards generic quantum many-body systems~\cite{Schollwock2011, RevModPhys.93.045003, RevModPhys.94.025005}.
It has been evolved into a new paradigm in modern many-body physics~\cite{Kuhn1962, Orus2014, Orus2019}.
TN methods hold great promise for tackling the challenging realm of strongly correlated quantum matters, such as cuprates and lattice quantum chromodynamics (QCD) with a finite density of baryons~\cite{WIESE2014246, NAGATA2022103991, Banuls2020}, which are typically inaccessible due to the notorious sign problem in Monte Carlo simulations.
As a special kind of one-dimensional TN, density matrix renormalization group (DMRG) has already achieved remarkable success in the investigation of $t$-$J$ model~\cite{SCHOLLWOCK201196, science.aal5304, PhysRevLett.127.097003, 2304.03963, 2311.15092,
PhysRevLett.127.097002, PhysRevB.108.054505, pnas.2109978118}.
2D fermionic TN studies on the square $t$-$J$ model also show the emergence of $d$-wave and many other competing states in its ground states subspace~\cite{PhysRevLett.113.046402, Yue2024, 2411.19218}.

In this work,
We use the imaginary-time evolutionary fermionic TN method to revisit the $t$-$J$ model on a square lattice directly in the thermodynamic limit and study another theoretically designed comparative model, highlighting the significant effects played by these nonlocal quantum phase strings.
The rest of this paper is organized as follows.
In Sec.~\ref{sigmatJ}, we introduce the quantum phase string effect in the $t$-$J$ and the comparative $\sigma{t}$-$J$ models. In Sec.~\ref{tensor}, we introduce the fermionic TN method.
In Sec.~\ref{results}, we use fermionic TN method to simulate $\sigma{t}$-$J$ model and show the physical effects played by the quantum phase string in their ground states.
In Sec.~\ref{sec:con_dis}, we make a brief conclusion and discussion.

\section{$\sigma t$-$J$ model and quantum phase string effect: A brief introduction}
\label{sigmatJ}

The $t$-$J$ model on a square lattice is widely believed to be able to capture the essence of a doped Mott insulator as the simplest model with spin SU(2) symmetry~\cite{PhysRevB.37.3759, RevModPhys.66.763, RevModPhys.78.17}. As a characteristic of strong correlation in such a model, doped holes moving in the AFM background will leave behind irreparable signs $(+)\times(-)\times(-)\dots$ known as phase strings~\cite{PhysRevLett.80.5401,PhysRevB.55.3894}.
Specifically, the partition function of the $t$-$J$ model can be explicitly expressed as~\cite{PhysRevB.77.155102}
\begin{equation}
    Z_{t\text{-}J}
    =\text{tr}\left(e^{-\beta{H}_{t\text{-}J}}\right)
    \equiv\sum_{C}\tau_{C}W[C]
\end{equation}
through a high-temperature expansion at any finite doping, where $\beta$ is the inverse temperature and $W[C]\geqslant 0$ denotes positive weights for any closed loop $C$ in the expansion. Here the Berry phase $\tau_{C}\equiv(-)^{N_{h}^{\downarrow}[C]+N_{\text{ex}}^{h}[C]}=\pm{1}$, representing an unconventional \emph{phase string sign} associated with the number of exchanges between holes and $\downarrow$-spins, $N_{h}^{\downarrow}[C] $, and a conventional Fermi-statistical sign associated with the number of hole-hole exchanges in loop $C$ as denoted by $N_{\text{ex}}^{h}[C]$. The singularity of the phase string $(-)^{N_{h}^{\downarrow}[C]}$ may be also seen by its equivalence to a mutual semionic statistics between charge and spin degrees of freedom upon doping ~\cite{PhysRevB.55.3894}.  

However, if we consider the so-called $\sigma{t}$-$J$ model $H_{\sigma{t}\text{-}J}\equiv{H}_{\sigma{t}}+H_{J}$, of which hopping term is modified to
\begin{equation}
    \begin{aligned}
    H_{\sigma{t}}
    &=-t\sum_{\langle{ij}\rangle,\sigma}\sigma\left(
        {c}_{i\sigma}^{\dagger}{c}_{j\sigma}+h.c.
    \right).
    \end{aligned}
    \label{eq:ham_sigmat}
\end{equation}
It can be proved that $Z_{\sigma{t}\text{-}J}=\sum_{C}(-)^{N_{\text{ex}}^{h}[C]}W[C]$, in which the quantum phase string associated with $N_{h}^{\downarrow}[C]$ is precisely "switched off", while the positive weight $W[C] $ remains the same. Then doped holes as fermions can propagate coherently without experiencing phase string frustrations.
This feature allows the $\sigma t$-$J$ model to be used as an ideal comparative benchmark of the quantum phase string effect in the $t$-$J$ model~\cite{Zhu2013,Zheng2018,PhysRevB.102.104512,PhysRevB.110.165127}. Note that the $\sigma{t}$-$J$ model is still not sign-problem free because of the residual Fermi-statistical sign $(-)^{N_{\text{ex}}^{h}[C]}$ in $Z_{\sigma{t}\text{-}J}$.
 
It is important to note that the no-double-occupancy constraint remains enforced in the $\sigma t$-$J$ model, indicating that this model still cannot be naively expected as a conventional doped semiconductor. This distinction is supported by previous DMRG studies~\cite{PhysRevB.110.165127}, which show that the observed Fermi surfaces for electrons with different spins exhibit the same volume $\delta \times A_{\mathrm{BZ}}$ (with $A_{\mathrm{BZ}}$ denoting the Brillouin zone area), which is different from the Luttinger sum rule for conventional Fermi liquids, where spinful electrons with density $\delta$ should give rise to a Fermi pocket with volume $\delta \times A_{\mathrm{BZ}}/2$.

To incorporate the Hilbert space constraint of $\sigma t$-$J$ model, we apply the parton construction in which the fermionic electron operator is expressed as $c_{i \sigma} = b_{i \sigma} f_i^{\dagger}$ , where  $b_{i \sigma}$  represents the bosonic spinon, and $f_i$  denotes the fermionic holon. At the mean field level (details provided in Appendix~\ref{App:mf}), we introduce the holon hopping $\hat{\kappa}_{i j}=f_i^{\dagger} f_j$, spinon hopping $\hat{\chi}_{i j, \sigma}=b_{i \sigma}^{\dagger} b_{j \sigma}$, and the RVB pairing $\hat{\Delta}_{i j}=\sum_\sigma \sigma b_{i, \sigma} b_{j,-\sigma}$ on nearest-neighbor bonds. All of these acquire finite expectation values under self-consistent mean-field calculations. As a result, spinons undergo Bose-Einstein condensation (BEC), and the two branches of gapless spinon modes with momentum difference $(\pi,\pi)$ combine to form the gauge-invariant spin operator $S^\pm$, indicating the magnetic long-range order in the $x$-$y$ plane.  In parallel, the fermionic holons form a well-defined Fermi surface, with spinon condensation, the identification  $c_{i \sigma} \sim f_i^{\dagger}$  indicates that the $\sigma t$-$J$ model exhibits Fermi liquid behavior along with magnetic order in $x$-$y$ plane.

Physically, the holon  motion in the $\sigma t$-$J$ model can disrupt the  magnetic order along the $z$ direction while preserving it in the $x$-$y$ plane. This is because the spinon hopping, $b_{i \uparrow}^{\dagger} b_{j \uparrow} - b_{i \downarrow}^{\dagger} b_{j \downarrow}$, contributes to the AFM correlations in the $x$-$y$ plane but favors ferromagnetic (FM) correlation along the $z$ plane. The former is compatible with the AFM correlations generated by RVB pairing $\hat \Delta_{ij}$, while the latter leads to frustration along the $z$ direction. Such contrasting behaviors of $S^\pm$ and $S^z$ basically manifest the explicit breaking of global SU(2) symmetry by the electron hopping term in Eq.~(\ref{eq:ham_sigmat}).

In the following, numerical results of the $\sigma t$-$J$ model  obtained using fermionic TN methods show strong consistency with the mean-field predictions discussed in this section, demonstrating that \emph{the parton construction formalism is particularly effective for handling the no-double-occupancy constraint when the phase string effect is removed}, as in the $\sigma t$-$J$ model.

\begin{figure}[tb]
    \centering
    \includegraphics[width=0.25\textwidth]{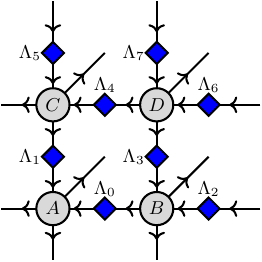}
    \caption{
        A $2 \times 2$ unit cell of a fTPS on a square lattice, containing four independent fermionic tensors $A, B, C, D$ on lattice sites, and eight independent fermionic Schmidt weight matrices $\Lambda_0, ..., \Lambda_7$ on nearest neighbor bonds.
        Indices of a tensor corresponding to super vector spaces (or dual spaces) are represented by outgoing (or incoming) arrows. 
    }
    \label{fig:fTPS}
\end{figure}

\section{Fermionic tensor product representation}
\label{tensor}
Ground states of these fermionic correlated models can be represented and approached by optimizing an infinite fermionic tensor product state (fTPS) (a.k.a. projected entangled pair state, PEPS).
Translational symmetry breaking may occur in the thermodynamic limit. 
As shown in Fig.~\ref{fig:fTPS}, we assume it has a $2 \times 2$ unit cell \cite{PhysRevB.81.165104, PhysRevB.88.115139} and can be written as:
\begin{equation}
    \Ket{\Psi}
    =\sum_{\{s\}}f\text{-tr}
    \left(T^{s_{0}}\Lambda\cdots\right)
    \Ket{\{s\}},
\end{equation}
"$f\text{-tr}$" means a fermionic contraction following the canonical $\mathbb{Z}_{2}$-graded tensor product isomorphism~\cite{Bultinck2018} for all internal bonds of the fTPS.
Each bond of a fermionic tensor further splits into two channels characterized by a $\mathbb{Z}_{2}^{f}$-fermion parity quantum number $m=0, 1$.
A rank-$5$ fermionic tensor is placed on each site \cite{PhysRevB.88.115139, PhysRevB.95.075108, Bultinck2018, PhysRevB.101.155105, Mortier2024} with four virtual bonds and one physical bond.
They are controlled by separated bond dimensions $D_{e}$ and $D_{o}$, respectively.
The total bond dimension is $D\equiv{D}_{e}+D_{o}$.
There is a Schmidt weight $\Lambda_0, ..., \Lambda_7$ on each bond, which is a rank-$2$ fermionic tensor (matrix).
Physical bonds are three dimensional, spanned by three orthogonal states
$\ket{0}$, 
$\ket{\uparrow} = c_{\uparrow}^{\dagger}\ket{0}$ and
$\ket{\downarrow} = c_{\downarrow}^{\dagger}\ket{0}$ constrained by the non-double occupancy in the $t$-$J$ model.

The Hamiltonian of these models should also be written as fermionic tensors in this representation (See more details in Appendix~\ref{App:tJ_rep} for the $t$-$J$ model).
From a randomly initialized state $\ket{\Psi}$, we use imaginary-time evolution methods to approach the ground state $\ket{\Psi_{0}}\propto \lim_{\tau\rightarrow\infty}e^{-\tau H} \ket{\Psi}$~\cite{2411.19218}.
Doping is controlled by tuning the chemical potential $\mu$ in a grand canonical ensemble.
We mainly focus on $t/J=3.0$ as experiments suggest for real materials~\cite{RevModPhys.78.17}.

\section{Numerical results}
\label{results}

We use variational uniform matrix product state (VUMPS) method~\cite{PhysRevB.97.045145, PhysRevB.98.235148, PhysRevB.105.195140, PhysRevB.108.035144} to measure physical quantities as targeting the boundary fixed points in the thermodynamic limit.
VUMPS is controlled by its own bond dimension $\chi$.
$\chi$ is gradually increased until convergence.
The magnetic order in the obtained state is measured by the staggered magnetization 
$
    M_i = \sqrt{
        \langle S_i^x \rangle^2
        + \langle S_i^y \rangle^2
        + \langle S_i^z \rangle^2
    }
$.
Real-space singlet pairing
$
\Delta_{ij}=\braket{c_{i\uparrow}c_{j\downarrow}-c_{i\downarrow}c_{j\uparrow}}/\sqrt{2}
$
can be also be directly measured for infinite systems.

\begin{figure}[t]
    \centering
    \includegraphics[width=0.35\textwidth]{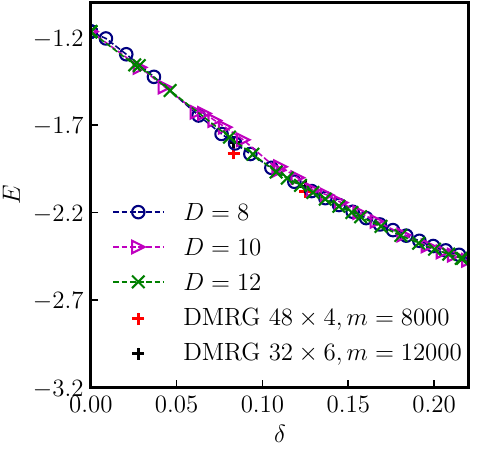}
    \caption{
    Ground state energies of the $\sigma{t}$-$J$ model.
    $m$ is the number of states kept in the DMRG sweeping.
    $t/J=3.0$.
    }
    \label{fig:sigmatJ_gs_ene}
\end{figure}

Firstly, we make some benchmark of the ground states energy of the $\sigma{t}$-$J$ model.
In Fig.~\ref{fig:sigmatJ_gs_ene} we can see that our infinite fTPS calculations
exhibit excellent energy convergence with that from finite-size $32\times 6$ cylinder DMRG~\cite{PhysRevB.110.165127}.
A $32\times 6$ cylinder is closer to a real 2D system than a $48\times4$ one.
In particular, For $D=12$, the extrapolated energy at half-filling is $E(\delta=0.0)\approx-1.165(7)$, a little higher by about $0.32\% $ than the best QMC result of $E_{0}=-1.1694$ for the Heisenberg model~\cite{PhysRevB.56.11678}.

\subsection{Compressibility}
\begin{figure}[t]
    \centering
    \includegraphics[width=0.35\textwidth]{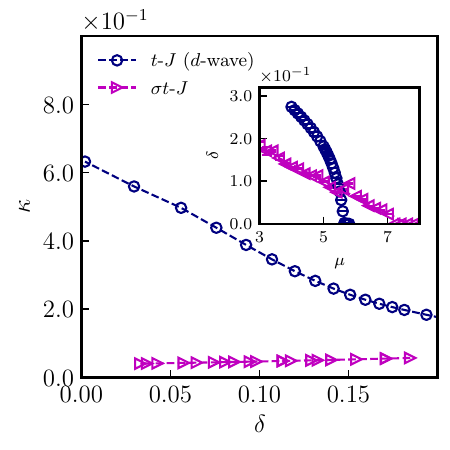}
    \caption{
    Compressibility $\kappa$ versus doping $\delta$ from the ground states of $t$-$J$ and $\sigma{t}$-$J$ models.
    Inset shows the doping $\delta$ versus the chemical potential $\mu$.
    $t/J=3.0$ and $D=12$.
    }
    \label{fig:tJ_sigmatJ_comp}
\end{figure}

Particularly, we are interested in the \emph{compressibility} of a quantum liquid, which is defined to be
\begin{equation}
    \kappa
    =-\frac{1}{V}\frac{\partial{V}}{\partial{P}}
    =\frac{1}{n^{2}}\frac{\partial{n}}{\partial\mu}.
\end{equation}
It characterizes the change of pressure with volume of a liquid.
In a standard Fermi liquid, we find that $\frac{\partial{n}}{\partial\mu}=\frac{N(0)}{1+F_{0}}\sim\text{const.}$, where $F_{0}$ is a Landau parameter and $N(0)$ is the quasiparticle density at the Fermi surface.
As shown in Fig.~\ref{fig:tJ_sigmatJ_comp}, the $\sigma{t}$-$J$ model fits well with the prediction of a Fermi liquid.
However, $t$-$J$ model exhibits a distinct non-Fermi liquid compressibility behavior especially at low doping levels, which signifies a collapse of Landau's quasiparticle picture in this regime.
Compressibility $\kappa$ sharply increases when approaching half-filling.
In the overdoped regime especially $\delta>20\%$, the $t$-$J$ model  progressively resembles a Fermi liquid and the compressibility of these two models becomes more and more identical, which meets our expectation in the phase diagram~\cite{Keimer2015} very well.
These results strongly suggest that quantum phase strings play a pivotal role in the non-Fermi liquid behavior of the $t$-$J$ model, particularly in the optimal and underdoped regimes.

\subsection{Magnetization}
\begin{figure}[t]
    \centering
    \includegraphics[width=0.52\textwidth]{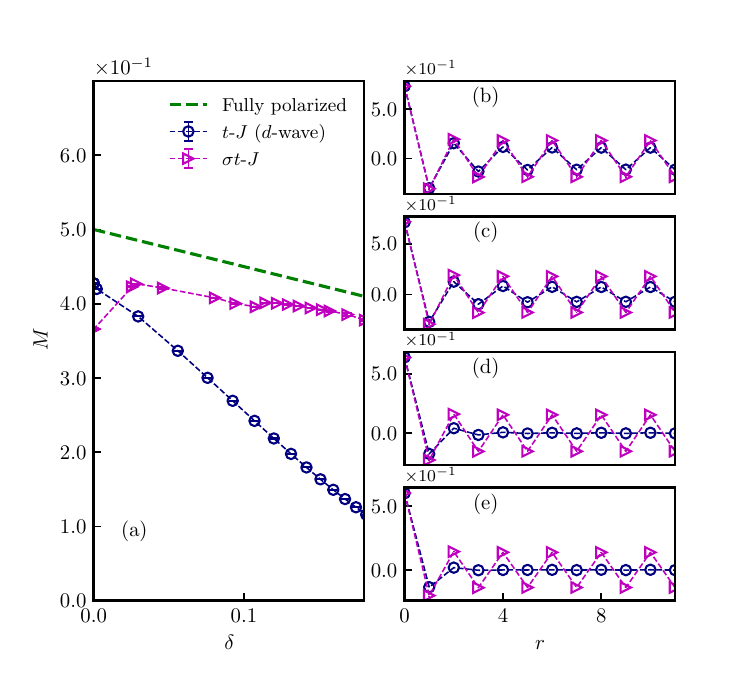}
    \caption{
    (a) Magnetizations in the ground states of $t$-$J$ and $\sigma{t}$-$J$ models. The dashed line shows the fully polarized magnetization $(1 - \delta)/2$. 
    (b, c, d, e) show spin-spin correlation functions $C_{s}(r)$ for these two models with doping levels around $\delta\approx3\%, 5\%, 15\%, 20\%$, respectively.
    $t/J=3.0$.
    $D=12$.
    }
    \label{fig:tJ_sigmatJ_mag}
\end{figure}

It is well-known that doped holes can suppress the AFM order rapidly in the phase diagram of cuprates~\cite{Keimer2015}.
After that, the superconductivity emerges.
As shown in Fig.~\ref{fig:tJ_sigmatJ_mag}(a), the magnetization of the $t$-$J$ model is suppressed much more quickly than that of the $\sigma{t}$-$J$ model.
This aligns well with experimental observations.
Meanwhile, the magnetization of the $\sigma{t}$-$J$ model keeps being relatively close to the fully polarized value $(1 - \delta) / 2$. 
The spin long-range AFM order is decoupled from the doped holes in the $\sigma{t}$-$J$ model.
To further illustrate their magnetic properties, we compute the spin-spin correlation function:
\begin{equation}
    C_{s}(r)\equiv\frac{1}{N}\sum_{(x,y)}\left\langle\mathbf{S}_{(x+r,y)}\cdot\mathbf{S}_{(x,y)}\right\rangle,
\end{equation}
where the summation of $(x, y)$ is over the whole unit cell and $N$ is the number of sites within the unit cell.
At small dopings with a strong AFM background persisting, both models exhibit similar pronounced oscillations in $C_{s}(r)$ as illustrated in Fig.~\ref{fig:tJ_sigmatJ_mag}(b).
However, at high doping levels, oscillations of $C_{s}(r)$ in the $t$-$J$ model gradually diminish (See Fig.~\ref{fig:tJ_sigmatJ_mag}(c, d, e)),
whereas they persist in the $\sigma{t}$-$J$ model across all doping levels, in agreement with the trends in Fig.~\ref{fig:tJ_sigmatJ_mag}(a).

\subsection{Superconductivity}
\begin{figure}[t]
    \centering
    \includegraphics[width=0.48\textwidth]{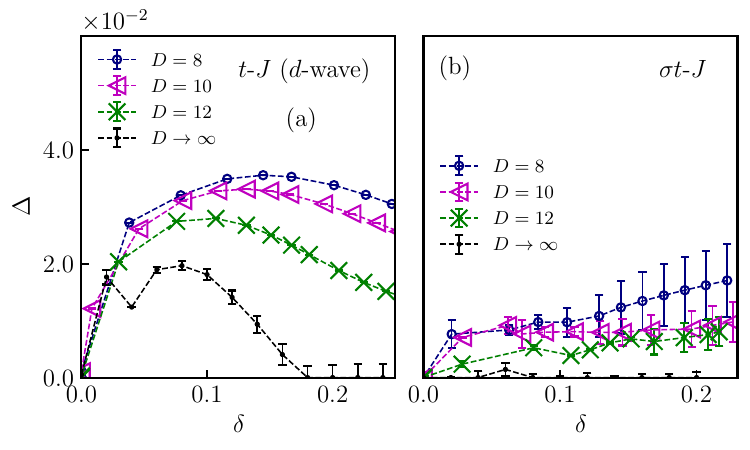}
    \caption{
    Superconductivity singlet pairing magnitudes in the (a) $t$-$J$ model (b) $\sigma{t}$-$J$ model.
    $t/J=3.0$.
    }
    \label{fig:tJ_sigmatJ_sc}
\end{figure}

As shown in Fig.~\ref{fig:tJ_sigmatJ_sc}, with finite $D$ in the TN, the superconducting singlet pairing amplitude $\Delta$ in the $t$-$J$ model is significantly enhanced compared to that in the $\sigma{t}$-$J$ model.
Remarkably, it does survive in the limit $D\rightarrow\infty$ when $\delta\lesssim 18\%$.
In contrast, the superconducting pairing amplitude in the $\sigma{t}$-$J$ model vanishes as $D\rightarrow\infty$ indicating there is no superconducting gap in the system.
These findings provide important insights into the high-$T_{c}$ pairing mechanism.
Note that P. W. Anderson pointed out that the high-$T_{c}$ pairing mechanism does originate from the AFM superexchange~\cite{science.235.4793.1196}.
Our results follow this idea and further elucidate that the primary pairing mechanism in the $t$-$J$ model should attribute to the quantum Berry phase strings associated with doped holes moving the AFM background.
It destroys the coherent Fermi liquid and open the unconventional superconducting gap.
The major pairing mechanism in a doped Mott insulator is dramatically different from the conventional Bardeen–Cooper–Schrieffer (BCS) electron-phonon coupling mechanism.
It can even protect the real-space electron pairing at elevated temperatures from the superconducting dome even up to the pseudogap phase~\cite{Zhou2019, PhysRevX.11.021054}.
Only the long-range coherence of superfluid is destroyed in the pseudogap phase and a possible PDW quantum liquid exists~\cite{2411.19218}.

\subsection{Single particle Green's function}
\begin{figure}[t]
    \centering
    \includegraphics[width=0.47\textwidth]{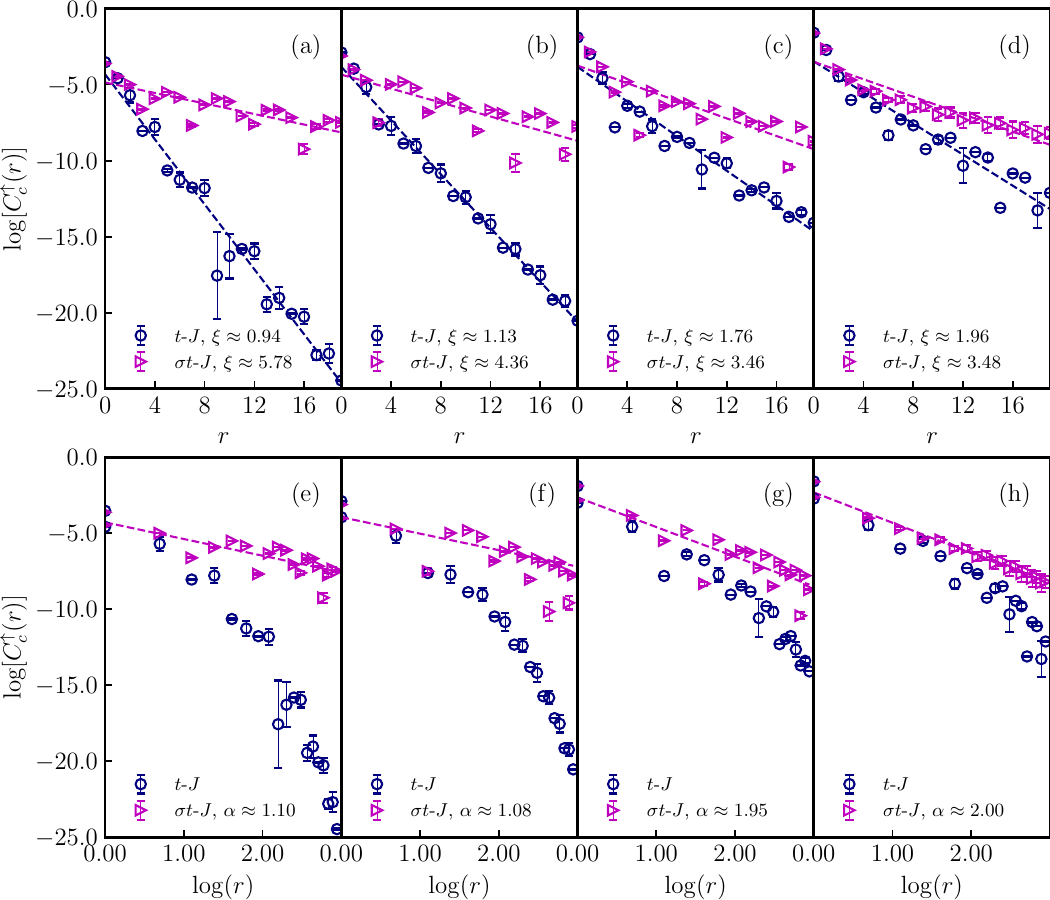}
    \caption{
    Single particle correlation functions $C_{c}^{\uparrow}(r)$ in the ground states of $t$-$J$ ($d$-wave) and $\sigma{t}$-$J$ models.
    (a, e), (b, f), (c, g) and (d, h) represent doping levels around $\delta\approx3\%, 5\%, 15\%, 20\%$ from left to right, respectively.
    (a, b, c, d) denote semi-logarithmic plots to extract the correlation length $\xi$.
    (e, f, g, h) denote double-logarithmic plots to extract the correlation power $\alpha$.
    $t/J=3.0$.
    $D=12$.
    }
    \label{fig:tJ_sigmatJ_cor_up}
\end{figure}

We further measure the correlation function of single electrons:
\begin{equation}
    C_{c}^{\sigma}(r)\equiv\frac{1}{N}\sum_{(x, y)}\left\langle{c}_{(x+r,y),\sigma}c_{(x,y),\sigma}^{\dagger}\right\rangle,
\end{equation}
which takes an asymptotic behavior as $C_{c}(r)\sim\exp\left(-r/\xi\right)/r^{\alpha}$.
Single particle Green's function $C_{c}(r)$ exhibits fundamentally different behavior between the $t$-$J$ model and the $\sigma{t}$-$J$ model in the underdoped regime, while gradually converging at higher dopings.
Representative doping levels around $\delta\approx 3\%, 5\%, 15\%, 20\%$ are plotted in Fig.~\ref{fig:tJ_sigmatJ_cor_up}.
In details, as shown in Fig.~\ref{fig:tJ_sigmatJ_cor_up}(a, b, c, d), the $t$-$J$ model displays a well-defined exponential decay behavior especially at low doping levels.
In the underdoped regime, the correlation length is very short ($\xi\approx{1}$).
While in the overdoped regime, the correlation length becomes slighter longer ($\xi\approx{2}$) implying the superconducting gap is diminishing.
Notably, the Green's functions in the $t$-$J$ model cannot be fitted by power-law decay (Fig.~\ref{fig:tJ_sigmatJ_cor_up}(e, f, g, h)).
Single particle in the $t$-$J$ model propagates very hard and decays very quickly.
In contrast, the $\sigma{t}$-$J$ model lacking the quantum phase string frustrations, displays significantly longer correlation and predominantly power-law behavior, reflecting a Fermi-liquid feature.
From underdoped to overdoped regimes, its correlation length becomes slightly shorter from $\xi\approx 6$ to $\approx 4$ and the power $\alpha$ is increased from $\approx 1$ to $\approx 2$.
This power-law behavior further corroborates the absence of a Cooper pairing gap in the $\sigma{t}$-$J$ model, as shown in Fig.~\ref{fig:tJ_sigmatJ_sc}.
Remarkably, both models show converging behavior at high dopings, consistent with the expected transition to a Fermi liquid state in the cuprate phase diagram at sufficiently large doping concentrations~\cite{Keimer2015}.
High dopings gradually submerges the phase string effect.
This doping-dependent evolution provides strong evidence for the crucial role of quantum phase strings in the underdoped and optimally doped regimes.

\begin{figure}[t]
    \centering
    \includegraphics[width=0.47\textwidth]{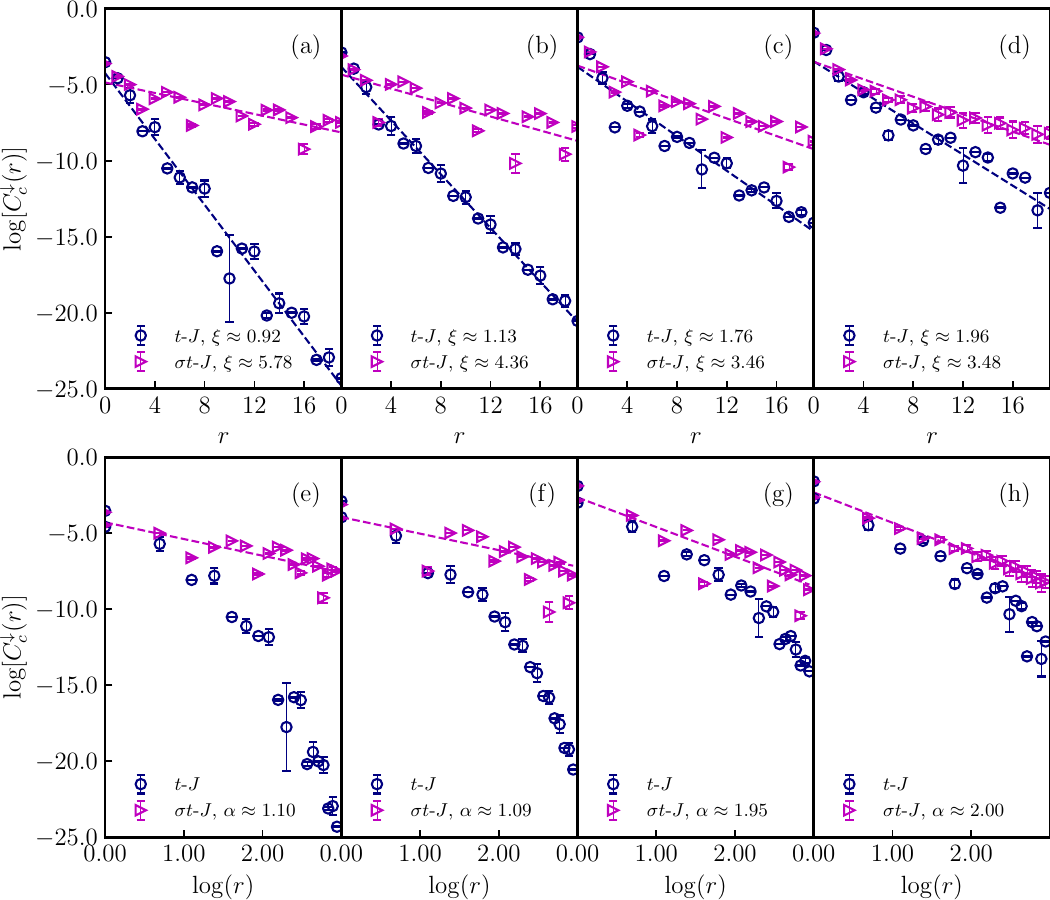}
    \caption{
    Single particle correlation functions $C_{c}^{\downarrow}(r)$ in the ground states of $t$-$J$ ($d$-wave) and $\sigma{t}$-$J$ models.
    (a, e), (b, f), (c, g) and (d, h) represent doping levels around $\delta\approx3\%, 5\%, 15\%, 20\%$ from left to right, respectively.
    (a, b, c, d) denote semi-logarithmic plots to extract the correlation length $\xi$.
    (e, f, g, h) denote double-logarithmic plots to extract the correlation power $\alpha$.
    $t/J=3.0$.
    $D=12$.
    }
    \label{fig:tJ_sigmatJ_cor_down}
\end{figure}

In Fig.~\ref{fig:tJ_sigmatJ_cor_down}, we show the Green's function for the branch spin-$\downarrow$, of which conclusion is exactly the same as spin-$\uparrow$ case.


\subsection{Next-nearest-neighbor (NNN) hopping}
\begin{figure}[t]
    \centering
    \includegraphics[width=0.48\textwidth]{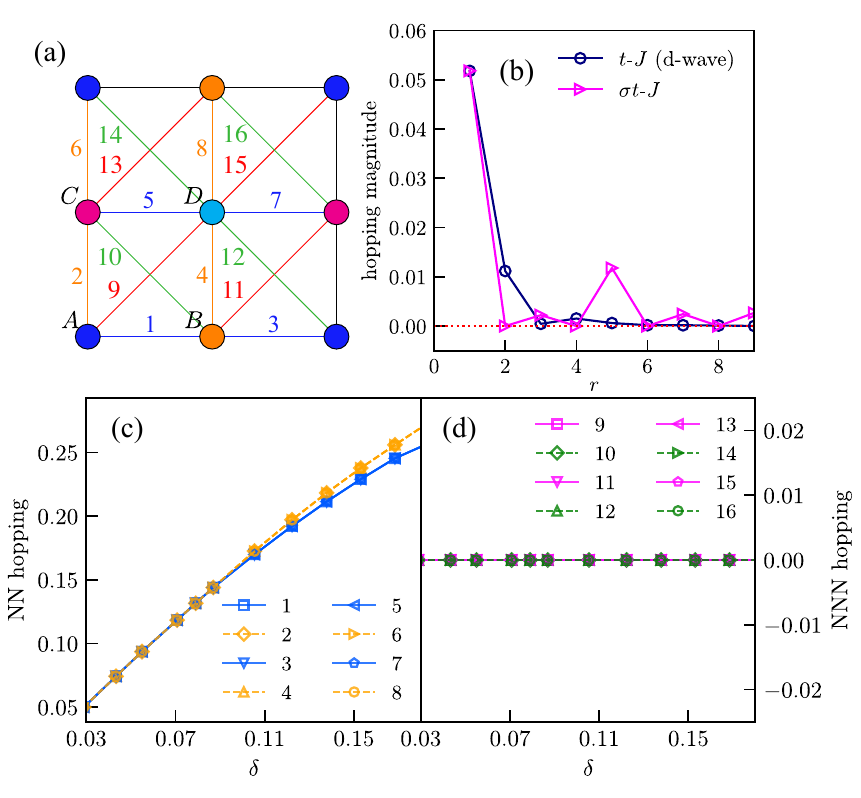}
    \caption{
    (a) Non-equivalent nearest neighbor (NN) bonds (1 to 8) and next nearest neighbor (NNN) bonds (9 to 16) in the $2 \times 2$ iPEPS unit cell. 
    (b-d) Magnitude of $\sigma t$-$J$ hopping term $\sum_{\sigma} \sigma \braket{c^\dagger_{i\sigma} c_{j\sigma} + h.c.}$ on (b) hotizontal bonds between two sites with distance $r$, (c) nearest neighbor (NN) bonds and (d) next nearest neighbor (NNN) bonds.
    In (b) the $\sigma t$-$J$ state is at $\delta \approx 6.3\%$, and we also compare with the usual $t$-$J$ hopping term in the d-wave state at $\delta \approx 7.7\%$. 
    (b) is measured for $D = 8$ states by CTMRG with boundary dimension $\chi = 32$, while (c-d) are measured for $D = 12$ with $\chi = 64$. 
    }
    \label{fig:hopping}
\end{figure}

The $\sigma t$-$J$ model hopping term $\sum_\sigma \sigma \braket{c^\dagger_{i\sigma} c_{j\sigma} + h.c.}$ also shows remarkable features that significantly distinct from the usual $t$-$J$ hopping term $\sum_\sigma \braket{c^\dagger_{i\sigma} c_{j\sigma} + h.c.}$. 
In particular, we observe that the $\sigma t$-$J$ hopping vanishes on the next nearest neighbor (NNN) bonds (Fig. ~\ref{fig:hopping}(d)), as well as horizontal and vertical bonds between two sites in the same sublattice (Fig. ~\ref{fig:hopping}(b)). Physically,  this is because the spinon hopping $b_{i \uparrow}^{\dagger} b_{j \uparrow}-b_{i \downarrow}^{\dagger} b_{j \downarrow}$, which contributes to AFM correlations in the $x$-$y$ plane on NNN bonds, is strongly suppressed in an AFM spin background where spins align parallel along the NNN bonds.  This indicates that the spinons form a specific resonating valence bond (RVB) state without same-sublattice hopping. These findings impose very strong constraints on the possible projective symmetry group (PSG) structure of the $\sigma t$-$J$ model, which we will explore further in future work.
In contrast, the $t$-$J$ hopping term is nonzero on same-sublattice bonds in the $d$-wave state \cite{2411.19218}, a feature that is critically important for the emergence of nodal quasiparticles. This stark difference between the two models suggests that the phase-string effect plays an essential role in the mechanism underlying emergent $d$-wave superconductivity in the $t$-$J$ model.

\section{Conclusion and discussion \label{sec:con_dis}}

In this work, by employing an advanced infinite 2D fermionic TN method, we have examined a novel quantum phase factor hidden in the $t$-$J$ model, known as the phase string. Physically, the phase string as a statistics-type Berry phase is equivalent to mutual semionic statistics between the spins and doped holes \cite{PhysRevB.55.3894}. It can be precisely turned off by inserting a spin-dependent sign $\sigma=\pm 1$ in the nearest-neighbor hopping term to result in the so-called $\sigma t$-$J$ model. It is important to note that both models have the same spin superexchange term, and are subjected to the same no-double-occupancy constraint, which are usually considered as the strong correlation characteristics in the $t$-$J$ model. However, by a comparative study based on the numerical fermionic TN approach, a sharp contrast in physical properties of two models has been revealed. The reduction of the AFM order by the increase of doping concentration found in the $t$-$J$ model is shown to level off in the $\sigma t$-$J$ model even though the $J$-term is the same.  A significant enhancement of the charge compressibility towards low-doping in the $t$-$J$ model is replaced by a flat Fermi liquid like one as a function of doping in the $\sigma t$-$J$ model. The single-particle propagator shows a substantial suppression in the former as compared to the Fermi liquid like power law decay in the latter. Correspondingly, a strong $d$-wave Cooper pairing of the doped holes exhibited in the former gets diminished in the latter.  


Therefore, the AFM spins and fermionic dopants are effectively decoupled in the $\sigma t$-$J$ model, with the doped holes forming a conventional Fermi liquid state of a small Fermi pocket like a doped semiconductor. Namely, it is clearly established that the phase string effect as the accumulation of the sign $\sigma=\pm 1$ in the hopping term of the $t$-$J$ model, leads to a new type of long-range many-body quantum entanglement between the spin and charge degrees of freedom~\cite{Zheng2018}. It is such a novel effect that is solely and systematically responsible for the suppression of the AFM long-range order by doping, the decoherence of the single-particle propagator with a much enhanced charge compressibility at low doping, and most importantly, a strong pairing in the Cooper channel which regains the quantum coherence by canceling the phase-string frustration. The accuracy of the present fermionic TN method may be continuously improved by further increasing the bond dimension $D$. For instance, the AFM order may be suppressed faster in the  $t$-$J$ model when the accuracy of the calculation is enhanced. Nonetheless, the stark distinction between the $t$-$J$ and $\sigma t$-$J$ models has already revealed an intriguing microscopic mechanism for strong correlation in doped Mott physics.      

Moreover, note that these unrenormalized quantum Berry phases are deeply connected to the numerical sign problem in a local spin-hole representation, representing a fundamental obstruction in certain nontrivial quantum many-body systems.
Conversely, strongly correlated quantum models such as the Heisenberg or $\sigma{t}$-$J$ model at half-filling is sign-problem free.
It can always be mapped to a classical system in a higher dimension using path-integral and efficiently simulated by statistical Monte Carlo sampling. Upon doping, even though there is still a residual fermionic sign problem associated with the doped holes in the $\sigma{t}$-$J$ model, the disappearance of the most nontrivial phase string sign structure in the ${t}$-$J$ model can result in a drastic change in physical consequences as revealed by the present TN simulations. We also propose that these quantum phase strings might be closely related to the highly competing ground states subspace of the $t$-$J$ model~\cite{PhysRevLett.113.046402,2411.19218}.
Further study on this is left for the future.


\begin{acknowledgements}

WZ would like to thank valuable discussions with Shuai A. Chen, Shou-Shu Gong, D. N. Sheng and Jia-Wei Mei.
This work is supported by funding from Hong Kong's Research Grants Council (CRF C7012-21GF and GRF No. 14302021).
J.X.Z was funded by the European Research Council (ERC) under the European Union’s Horizon 2020 research and innovation program (Grant Agreement No. 853116, acronym TRANSPORT), and was also supported in part by grant NSF PHY-2309135 to the Kavli Institute for Theoretical Physics (KITP).
The financial support (Z.Y.W) by MOST of China (Grant No.~2021YFA1402101) and NSF of China (Grant No.~12347107) is also acknowledged. 

\end{acknowledgements}

\appendix

\section{Mean-Field parton construction  of $\sigma{t}$-$J$ model}\label{App:mf}
To describe the low-lying behavior of the $\sigma{t}$-$J$ model, we employ the mean-field parton construction approach, where the fermionic electron operator is expressed as:
\begin{equation}\label{cbf}
    c_{i \sigma}=b_{i \sigma} f_i^{\dagger},
\end{equation}
where  $b_{i\sigma}$  represents the bosonic spinon and  $f_i^{\dagger}$  denotes the fermionic holon. Using this representation, the $\sigma t$-$J$ model can be reformulated as:
\begin{eqnarray}\label{fHstJ}
    H_{\sigma t \text{-}J}=t \sum_{\langle i j\rangle, \sigma} \sigma \hat{\chi}_{i j, \sigma} \hat{\kappa}_{i, j}^{\dagger}+\text{H.c.}-\frac{J}{2} \sum_{i j} \hat{\Delta}_{i j}^{\dagger} \hat{\Delta}_{ij}
\end{eqnarray}
where 
\begin{eqnarray}
    \hat{\kappa}_{ij} &=& f_{i}^{\dagger} f_j\\
    \hat{\chi}_{ij, \sigma} &=& b_{i \sigma}^{\dagger} b_{j \sigma}\\
\hat{\Delta}_{ij}&=&\sum_\sigma \sigma b_{i, \sigma} b_{j,-\sigma}
\end{eqnarray}
Using the mean-field ansatz: $\hat{\kappa}_{ij}= \text{i} \kappa$, $\hat{\chi}_{ij,\sigma}= \text{i} \sigma \chi$, and $\hat{\Delta}_{ij}= \text{i} \Delta$,
the Hamiltonian at the mean-field level is given by $H_f + H_b$, where the holon part is:
\begin{eqnarray}\label{Hf}
    \begin{aligned}
    H_f =&  -4 t \chi\sum_k \mathcal{E}_k f_k^{\dagger}f_k -\mu \sum_k\left(f_k^{\dagger} f_k-\delta\right)
    \end{aligned}
\end{eqnarray}
and the spinon part is:
\begin{eqnarray}\label{Hb}
    H_b =&-2t\kappa \sum_{k, \sigma} \sigma   \mathcal{E}_k b_{k \sigma}^{\dagger} b_{k \sigma} + J \Delta \sum_k \mathcal{E}_k \left(b_{-k \downarrow} b_{k \uparrow}+\mathrm{H.c.}\right)\nonumber \\ 
    &+J \Delta^2 N -8 t N \chi \kappa + \lambda \sum_k \left(\sum_{\sigma} b_{k \sigma}^{\dagger} b_{k \sigma} -1+\delta\right)\nonumber\\
\end{eqnarray}
where  $\mathcal{E}_k=\sin k_x+\sin k_y$, and $\kappa$, $\chi$, $\Delta$  are real parameters. Here, $\mu$ and $\lambda$ are the chemical potentials to control the holon number $\delta$ and spinon number $1-\delta$, respectively. The values of these mean-field parameters can be determined via self-consistent calculations. 

\begin{figure}[t]
    \centering
    \includegraphics[width=0.7\linewidth]{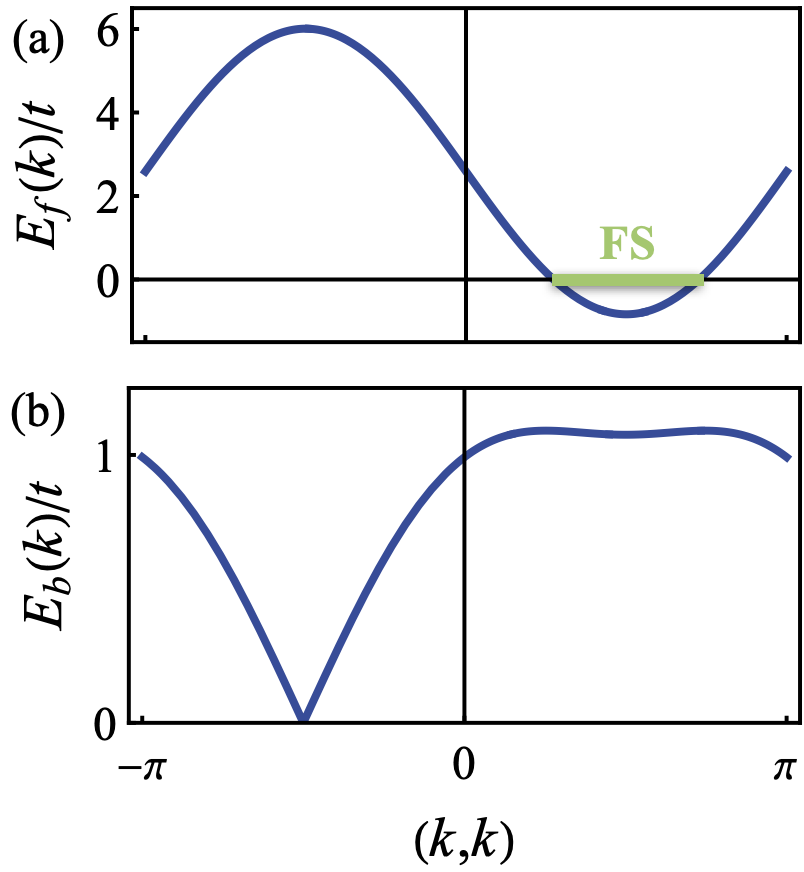}
    \caption{
    Using the parameters obtained from the self-consistent equations for the case of $t/J = 3$ and doping density $\delta = 1/12$, the dispersions for holons, $E_f(k)$ in Eq. \ref{Efk}, and spinons, $E_b(k)$ in Eq. \ref{Ebk}, are shown in (a) and (b). The Fermi surface (FS) of holons is labeled by the thick green line.
    }
    \label{fig:MFT}
\end{figure}

From Eq.~(\ref{Hf}), the dispersion for holon is given by 
\begin{equation}\label{Efk}
    E_f(k)=-4 t \chi \mathcal{E}_k-\mu,
\end{equation}
which is shown in Fig.~\ref{fig:MFT}(a) based on the calculated mean-field parameters, with the Fermi surface indicated by the thick green line. Similarly, from Eq.~\ref{Hb}, after applying the following Bogoliubov transformation, we obtain:
\begin{equation}
    H_b=\sum_{k,\sigma} E_b(k) \gamma_{k,\sigma}^\dagger \gamma_{k,\sigma}
\end{equation}
with the dispersion 
\begin{equation}\label{Ebk}
    E_b(k) = \sqrt{\left(\lambda-2 t \kappa \mathcal{E}_k\right)^2-J^2 \Delta^2 \mathcal{E}_k^2}
\end{equation}
for the Bogoliubov spinon $\gamma_{k\uparrow}= u_k b_{k\uparrow}+ v_k b_{-k\downarrow}^\dagger$ and $\gamma_{k\downarrow}= u_k b_{-k\downarrow}+ v_k b_{k\uparrow}^\dagger$. The dispersion $E_b$ is shown in Fig.~\ref{fig:MFT}(b), which exhibits condensation with gapless excitation near $(-\pi/2,-\pi/2)$. As the spinons condense, spin excitations become relevant at low energies. The low-lying spinon modes are located at momenta $(\sigma \pi/2, \sigma \pi/2)$, tso the relevant modes near the two valleys can be organized as $\psi=\left[b_{+\downarrow},-b_{+\uparrow}^{\dagger}, b_{-\uparrow}, b_{-\downarrow}^{\dagger}\right]^T$, where $\pm$ label the two valleys at momenta $\pm( \pi/2, \pi/2)$. Considering bilinear, gauge-invariant operators of the form, $\psi^\dagger \Gamma \psi $, we find that that the  AFM spin operator $S^+_{\boldsymbol{Q}} \sim b_{-\uparrow}^\dagger b_{+\downarrow}$ and $S^-_{\boldsymbol{Q}} \sim b_{+\downarrow}^\dagger b_{-\uparrow}$ can be constructed, but the the $z$-component  $S^z_{\boldsymbol{Q}}\sim\sum_{\sigma}b_{+\sigma}^\dagger b_{-\sigma}$ is not supported in this basis. This implies that spinon condensation leads to AFM ordering in the $x$-$y$ plane.

It is important to note that the condensation of the $b$-spinon implies the identification $c_{i \sigma} \sim f_i^{\dagger}$, meaning that the behavior of electrons mirrors that of the fermionic holons, which exhibit a Fermi surface. As a result, the parton construction at the mean-field level indicates that the $\sigma t$-$J$ model exhibits a confined phase, characterized by a Fermi liquid coexisting with magnetic order. Consequently, gauge fluctuation effects, which are typically relevant in parton constructions involving deconfined particles, do not need to be considered in this context.

\section{Notes on the TN method for fermions \label{sec:tn_fermions}}
\subsection{Super vector space and graded tensor product}

To accurately describe the nature of ferminons in the framework of TN, two essential theoretical considerations are required.
First, we need to incorporate the generic $\mathbb{Z}_{2}$ parity symmetry that is conserved by fermionic quantum many-body systems.
This means each tensor must fulfill the Abelian fusion rule of $\mathbb{Z}_{2}$ charges, meaning that the total charges associated with the outgoing bonds must be equal to the total charges associated with the incoming bonds in the sense of the modular two.
That is, the parity for each tensor must be \emph{even}.
Therefore, if a many-body wavefunction is constructed by these $\mathbb{Z}_{2}$ symmetric tensors under periodic boundary conditions, its parity, namely the total parity summed over from all physical bonds, must be \emph{even} since all internal bonds are contracted.
The odd parity sector indeed can be constructed from the even one, by such as inserting some kind of fluxes or imposing anti-periodic boundary conditions.
It is worth to mention that we can easily construct wavefunctions with any parity under open boundary conditions since we can freely add virtual bonds and pin down their quantum numbers on the boundary~\cite{Rakov2018}.

Secondly, we need to encode the Fermi statistics into the TN, mathematically achieved by the \emph{graded tensor product structure}.
A $\mathbb{Z}_{2}$-graded \emph{super vector space} is the direct sum of two ordinary vector spaces~\cite{PhysRevB.95.075108, Bultinck2018, PhysRevB.101.155105} as
\begin{equation}
    V=V^{[0]}\oplus{V^{[1]}}.
\end{equation}
$V^{[0]}$ and $V^{[1]}$ represent the even and odd subspaces, respectively.
This structure allows us to distinguish between even and odd elements, which is crucial for correctly representing the antisymmetric nature of fermions.
A basis for $V$ written as $\vert\mathbf{e}_{i}^{m}\rangle$ should be denoted which subspace it belongs to by $m=0, 1$.
$i$ is the normal index running over the dimension of this subspace.
Any vector having a definite parity is called \emph{homogenuous}.
The dual space $V^{*}$ spanned by $\left\{\langle\mathbf{e}_{j}^{n}\vert\right\}$ satisfying the orthogonality $\langle\mathbf{e}_{j}^{n}\vert\mathbf{e}_{i}^{m}\rangle=\delta_{ij}\delta_{mn}$.
The tensor product of these kinds of super vector spaces are generalized and called $\mathbb{Z}_{2}$-graded.
For example,
\begin{equation}
    \vert\mathbf{e}_{i}^{m}\rangle\otimes_{g}\vert\mathbf{e}_{j}^{n}\rangle\in{V}_{0}\otimes_{g}V_{1},
\end{equation}
where $\vert\mathbf{e}_{i}^{m}\rangle\in{V}_{0}$ and $\vert\mathbf{e}_{j}^{n}\rangle\in{V}_{1}$.
A \emph{canonical $\mathbb{Z}_{2}$-graded tensor product isomorphism} is defined to be 
\begin{equation}
    \mathcal{F}:
    \vert\mathbf{e}_{i}^{m}\rangle\otimes_{g}\vert\mathbf{e}_{j}^{n}\rangle
    \rightarrow
    (-)^{mn}\vert\mathbf{e}_{j}^{n}\rangle\otimes_{g}\vert\mathbf{e}_{i}^{m}\rangle.
\end{equation}
Physically, $\mathcal{F}$ refers to reordering fermions and defines how to permute the bonds of a $\mathbb{Z}_{2}$-graded tensor.
A rank-$\left(r, s\right)$ $\mathbb{Z}_{2}$-graded tensor can be regarded as a linear map from $r$ dual super vector spaces to $s$ super vector spaces.
In practice, we record a \emph{dual} quantity associated to each $\mathbb{Z}_{2}$-graded tensor representing the types of vector space on each bond.
For example,
\begin{equation}
    \mathsf{T}
    =T_{j_{0}j_{1}j_{2}}^{n_{0}n_{1}n_{2}}\vert\mathbf{e}_{j_{0}}^{n_{0}}\rangle\otimes_{g}\vert\mathbf{e}_{j_{1}}^{n_{1}}\rangle\otimes_{g}\langle\mathbf{e}_{j_{2}}^{n_{2}}\vert
\end{equation}
is a rank-$\left(2, 1\right)$ and dual-$(0, 0, 1)$ $\mathbb{Z}_{2}$-graded tensor belonging to ${V}_{0}\otimes_{g}V_{1}\otimes{V}_{2}^{*}$. $\mathsf{T}$ has a definite parity denoted by $\vert\mathsf{T}\vert\equiv{n_{0}+n_{1}+n_{2}}$ mod $2$.
$T_{j_{0}j_{1}j_{2}}^{n_{0}n_{1}n_{2}}\in\mathbb{C}$ is an ordinary complex degeneracy tensor.
The upper indices $n_{0, 1, 2}=0, 1$ are also called \emph{fermionic indices}.

If the $\mathbb{Z}_{2}$-parity symmetry is preserved, a $\mathbb{Z}_{2}$-graded tensor can be further promoted to be $\mathbb{Z}_{2}$-symmetric.
Each block should be $\mathbb{Z}_{2}$-symmetric.

\subsection{Write $t$-$J$ model in the fermionic tensor representation\label{App:tJ_rep}}

In this section, we explain in detail how to express the Hamiltonian of $t$-$J$ model Eq.~\eqref{eq:ham_tJ} and related operators in the representation of $\mathbb{Z}_{2}$-graded fermionic tensors.
In the first place, we know that local spin operators can be written in terms of fermions as
\begin{equation}
    S_{j}^{a}
    =\frac{1}{2}F_{j}^{\dagger}\sigma^{a}F_{j}, \quad F_{j}=\left(c_{j\uparrow}, c_{j\downarrow}\right)^{T}.
    \label{eq:spin_op_fermion}
\end{equation}
$\sigma^{a}, a=x, y, z$ are the Pauli matrices.
Note that, for the $t$-$J$ model, the one-site double occupancy is strictly projected out therefore the local Hilbert space only consists of three basis vectors, which can be defined as
\begin{equation}
    \ket{\mathbf{e}_0^0}=\vert{0}\rangle, \quad
    \ket{\mathbf{e}_0^1}=c_{\uparrow}^{\dagger}\vert{0}\rangle, \quad
    \ket{\mathbf{e}_1^1}=c_{\downarrow}^{\dagger}\vert{0}\rangle
\end{equation}
in the $\mathbb{Z}_{2}$-graded super vector space using fermionic operators. The upper script labels the fermion parity quantum number, and the lower script labels the basis vectors in the subspace given the parity. 
The double occupied state $\ket{\mathbf{e}_{1}^{0}}$ should be projected out and the closure identity condition is reduced to
\begin{equation}
    \mathbbm{1}=\ket{\mathbf{e}_0^0}\bra{\mathbf{e}_0^0}+\ket{\mathbf{e}_0^1}\bra{\mathbf{e}_0^1}+\ket{\mathbf{e}_1^1}\bra{\mathbf{e}_1^1}.
    \label{eq:tJ_fermion_identity}
\end{equation}
Single-body fermionic operators in this projected space can be written as fermionic tensors:
\begin{equation}
    \begin{aligned}
        c_{\downarrow}^{\dagger}
        &=\ket{\mathbf{e}_1^1}\bra{\mathbf{e}_0^0}, \quad
        c_{\downarrow}
        =\ket{\mathbf{e}_0^0}\bra{\mathbf{e}_1^1}, \\
        c_{\uparrow}^{\dagger} 
        &=\ket{\mathbf{e}_0^1}\bra{\mathbf{e}_0^0}, \quad
        c_{\uparrow}
        =\ket{\mathbf{e}_0^0}\bra{\mathbf{e}_0^1}.
    \end{aligned}
    \label{eq:tJ_cfermions_graded_tensor}
\end{equation}
Spin operators are
\begin{equation}
    \begin{aligned}
    S^{x}
    &=\frac{1}{2}\left(\ket{\mathbf{e}_0^1}\bra{\mathbf{e}_1^1}+\ket{\mathbf{e}_1^1}\bra{\mathbf{e}_0^1}\right), \\
    S^{y}
    &=-\frac{\text{i}}{2}\left(\ket{\mathbf{e}_0^1}\bra{\mathbf{e}_1^1}-\ket{\mathbf{e}_1^1}\bra{\mathbf{e}_0^1}\right), \\
    S^{z}
    &=\frac{1}{2}\left(\ket{\mathbf{e}_0^1}\bra{\mathbf{e}_0^1}-\ket{\mathbf{e}_1^1}\bra{\mathbf{e}_1^1}\right).
    \end{aligned}
\end{equation}
Our task is to write a two-body gate Hamiltonian of $t$-$J$ model into a fermionic tensor
\begin{equation}
    \mathsf{H}
    =H_{i^{\prime}j^{\prime}ij}^{n_{i^{\prime}}n_{j^{\prime}}n_{i}n_{j}}\vert\mathbf{e}_{i^{\prime}}^{n_{i^{\prime}}}\rangle\vert\mathbf{e}_{j^{\prime}}^{n_{j^{\prime}}}\rangle\langle\mathbf{e}_{i}^{n_{i}}\vert\langle\mathbf{e}_{j}^{n_{j}}\vert
\end{equation}
regardless of specific parameters.
Firstly, to avoid possible extra signs, we need to write the two-body gate Hamiltonian tensor in the \emph{natural tensor product order} $H_{i^{\prime}ij^{\prime}j}^{n_{i^{\prime}}n_{i}n_{j^{\prime}}n_{j}}$ as placing the tensor product site by site.
The hopping terms
\begin{equation}
    \begin{aligned}
        c_{i\uparrow}^{\dagger}c_{j\uparrow}
        &= \left(\ket{\mathbf{e}_0^1}\bra{\mathbf{e}_0^0}\right)_{i}\otimes_{g}\left(\ket{\mathbf{e}_0^0}\bra{\mathbf{e}_0^1}\right)_{j}, \\
        c_{i\uparrow}c_{j\uparrow}^{\dagger}
        &=\left(\ket{\mathbf{e}_0^0}\bra{\mathbf{e}_0^1}\right)_{i}\otimes_{g}\left(\ket{\mathbf{e}_0^1}\bra{\mathbf{e}_0^0}\right)_{j}, \\
        c_{i\downarrow}^{\dagger}c_{j\downarrow}
        &=\left(\ket{\mathbf{e}_1^1}\bra{\mathbf{e}_0^0}\right)_{i}\otimes_{g}\left(\ket{\mathbf{e}_0^0}\bra{\mathbf{e}_1^1}\right)_{j}, \\
        c_{i\downarrow}c_{j\downarrow}^{\dagger}
        &=\left(\ket{\mathbf{e}_0^0}\bra{\mathbf{e}_1^1}\right)_{i}\otimes_{g}\left(\ket{\mathbf{e}_1^1}\bra{\mathbf{e}_0^0}\right)_{j}
    \end{aligned}
\end{equation}
lead to the nonzero tensor elements $H_{1001}^{1001}=H_{0000}^{1001}=1, H_{0110}^{0110}=H_{0000}^{0110}=-1$.
Note that there is a minus sign in the Hermitian conjugated hopping term.
We can also write down the superconductivity order operators:
\begin{equation}
    \begin{aligned}
    c_{i\uparrow}c_{j\uparrow}
    &=\left(\ket{\mathbf{e}_0^0}\bra{\mathbf{e}_0^1}\right)_{i}\otimes_{g}\left(\ket{\mathbf{e}_0^0}\bra{\mathbf{e}_0^1}\right)_{j}, \\
    c_{i\downarrow}c_{j\downarrow}
    &=\left(\ket{\mathbf{e}_0^0}\bra{\mathbf{e}_1^1}\right)_{i}\otimes_{g}\left(\ket{\mathbf{e}_0^0}\bra{\mathbf{e}_1^1}\right)_{j}.
    \end{aligned}
\end{equation}
For the Heisenberg term, we rewrite it to
\begin{equation}
    \begin{aligned}
        &\mathbf{S}_{i}\cdot\mathbf{S}_{j}-\frac{n_{i}n_{j}}{4} \\
        &=\frac{1}{2}\left({S}_{i}^{+}S_{j}^{-}+S_{i}^{-}S_{j}^{+}\right)+S_{i}^{z}S_{j}^{z}-\frac{n_{i}n_{j}}{4} \\
        &=\frac{1}{2}\left(c_{i\uparrow}^{\dagger}c_{i\downarrow}c_{j\downarrow}^{\dagger}c_{j\uparrow}+c_{i\downarrow}^{\dagger}c_{i\uparrow}c_{j\uparrow}^{\dagger}c_{j\downarrow}\right)-\frac{1}{2}\left(n_{i\uparrow}n_{j\downarrow}+n_{i\downarrow}n_{j\uparrow}\right) \\
        &=-\frac{1}{2}\sum_{\sigma,\sigma^{\prime}}\sigma\sigma^{\prime}c_{i\sigma}^{\dagger}c_{i\sigma^{\prime}}c_{j,-\sigma}^{\dagger}c_{j,-\sigma^{\prime}}.
    \end{aligned}
\end{equation}
Along with
\begin{equation}
    \begin{aligned}
        c_{\downarrow}^{\dagger}c_{\uparrow}
        &=\ket{\mathbf{e}_1^1}\bra{\mathbf{e}_0^1}, \quad
        c_{\uparrow}^{\dagger}c_{\downarrow}
        =\ket{\mathbf{e}_0^1}\bra{\mathbf{e}_1^1}, \\
        n_{\downarrow}
        &=c_{\downarrow}^{\dagger}c_{\downarrow}
        =\ket{\mathbf{e}_1^1}\bra{\mathbf{e}_1^1}, \quad
        n_{\uparrow}
        =c_{\uparrow}^{\dagger}c_{\uparrow}
        =\ket{\mathbf{e}_0^1}\bra{\mathbf{e}_0^1},
    \end{aligned}
\end{equation}
we have
\begin{equation}
    \begin{aligned}
        c_{i\downarrow}^{\dagger}c_{i\uparrow}c_{j\uparrow}^{\dagger}c_{j\downarrow}
        &=\left(\ket{\mathbf{e}_1^1}\bra{\mathbf{e}_0^1}\right)_{i}\otimes_{g}\left(\ket{\mathbf{e}_0^1}\bra{\mathbf{e}_1^1}\right)_{j}, \\
        c_{i\uparrow}^{\dagger}c_{i\downarrow}c_{j\downarrow}^{\dagger}c_{j\uparrow}
        &=\left(\ket{\mathbf{e}_0^1}\bra{\mathbf{e}_1^1}\right)_{i}\otimes_{g}\left(\ket{\mathbf{e}_1^1}\bra{\mathbf{e}_0^1}\right)_{j}, \\
        n_{i\downarrow}n_{j\uparrow}
        &=\left(\ket{\mathbf{e}_1^1}\bra{\mathbf{e}_1^1}\right)_{i}\otimes_{g}\left(\ket{\mathbf{e}_0^1}\bra{\mathbf{e}_0^1}\right)_{j}, \\
        n_{i\uparrow}n_{j\downarrow}
        &=\left(\ket{\mathbf{e}_0^1}\bra{\mathbf{e}_0^1}\right)_{i}\otimes_{g}\left(\ket{\mathbf{e}_1^1}\bra{\mathbf{e}_1^1}\right)_{j}.
    \end{aligned}
    \label{eq:tj_heisenberg_z2gtensors}
\end{equation}
We realize that the first and second terms in Eq.~\eqref{eq:tj_heisenberg_z2gtensors} are off-diagonal and the third and fourth ones are diagonal.
All these non-vanishing tensor elements are $H_{1001}^{1111}=H_{0110}^{1111}=\frac{1}{2}, H_{1100}^{1111}=H_{0011}^{1111}=-\frac{1}{2}$.

For the chemical potential diagonal term, on a infinite square lattice, we can rewrite it in a symmetric form as
\begin{equation}
    \sum_{j}n_{j}
    =\frac{1}{4}\sum_{\langle{ij}\rangle}\left(n_{i}+n_{j}\right)
\end{equation}
since the number of bonds is double of the number of sites.
Thereby we have
\begin{equation}
    \begin{aligned}
        n_{i}
        &=\left(n_{\downarrow}+n_{\uparrow}\right)_{i}\otimes_{g}\mathbbm{1}_{j} \\
        &=\left(\ket{\mathbf{e}_1^1}\bra{\mathbf{e}_1^1}+\ket{\mathbf{e}_0^1}\bra{\mathbf{e}_0^1}\right)_{i}\otimes_{g}\left(\sum_{k, n}\vert\mathbf{e}_{k}^{n}\rangle\langle\mathbf{e}_{k}^{n}\vert\right)_{j}.
    \end{aligned}
\end{equation}
Note that here we only have six non-vanishing tensor elements because of the closure condition Eq.~\eqref{eq:tJ_fermion_identity} rather than eight.
Similarly,
\begin{equation}
    \begin{aligned}
        n_{j}
        &=\mathbbm{1}_{i}\otimes_{g}\left(n_{\uparrow}+n_{\downarrow}\right)_{j} \\
        &=\left(\sum_{k, n}\vert\mathbf{e}_{k}^{n}\rangle\langle\mathbf{e}_{k}^{n}\vert\right)_{i}\otimes_{g}\left(\ket{\mathbf{e}_1^1}\bra{\mathbf{e}_1^1}+\ket{\mathbf{e}_0^1}\bra{\mathbf{e}_0^1}\right)_{j}
    \end{aligned} 
\end{equation}
and all these non-vanishing items should be accumulated.
After building $H_{i^{\prime}ij^{\prime}j}^{n_{i}^{\prime}n_{i}n_{j}^{\prime}n_{j}}$, we permute it to a new order $H_{i^{\prime}j^{\prime}ij}^{n_{i}^{\prime}n_{j}^{\prime}n_{i}n_{j}}$, which is more convenient to be used in the practical numerical simulations.
Some extra signs may arise according to the fermionic rule.
We can find that the parity of $\mathsf{H}$ is even thus it is $\mathbb{Z}_{2}$-symmetric as we expected.

The onsite particle number operator is
\begin{equation}
    \mathsf{N}
    =N_{i^{\prime}i}^{n_{i^{\prime}}n_{i}}\vert\mathbf{e}_{i^{\prime}}^{n_{i^{\prime}}}\rangle\langle\mathbf{e}_{i}^{n_{i}}\vert.
\end{equation}
Since $c_{\uparrow}^{\dagger}c_{\uparrow}=\ket{\mathbf{e}_0^1}\bra{\mathbf{e}_0^1}$ and $c_{\downarrow}^{\dagger}c_{\downarrow}=\ket{\mathbf{e}_1^1}\bra{\mathbf{e}_1^1}$, we have $N_{00}^{11}=N_{11}^{11}=1$.
Others are zero.

The magnetization operator is
\begin{equation}
    \mathsf{S}^{z}
    =S_{i^{\prime}i}^{n_{i^{\prime}}n_{i}}\vert\mathbf{e}_{i^{\prime}}^{n_{i^{\prime}}}\rangle\langle\mathbf{e}_{i}^{n_{i}}\vert.
\end{equation}
Since $S^{z}=\frac{1}{2}\left(c_{\uparrow}^{\dagger}c_{\uparrow}-c_{\downarrow}^{\dagger}c_{\downarrow}\right)$, we have $S_{00}^{11}=\frac{1}{2}, S_{11}^{11}=-\frac{1}{2}$.
Others are zero.

\subsection{Isometry, graded conjugation and unitarity of fermionic tensors}

A one-body identity fermionic tensor (matrix) is written as
\begin{equation}
    \mathsf{I}
    =\sum_{j, n}\vert\mathbf{e}_{j}^{n}\rangle\langle\mathbf{e}_{j}^{n}\vert,
    \label{eq:gtensor_id}
\end{equation}
which is a rank-(1, 1) and dual-$(0, 1)$.
It has two sectors represented by $\mathbbm{Z}_{2}$-fermionic quantum numbers $\mathbf{n}=(0, 0)$ and $(1, 1)$.
Each block is comprised of an ordinary identity matrix $\mathbbm{1}$, respectively.
On the other hand, we could permute $\mathsf{I}$ to a dual-$(1, 0)$ identity operator denoted by $\bar{\mathsf{I}}$, in which the sector with $\mathbf{n}=(1, 1)$ is $-\mathbbm{1}$ due to the fermionic graded tensor product isomorphism.
We call $\bar{\mathsf{I}}$ as a \emph{super identity} of fermionic tensor and regard $\bar{\mathsf{I}}\cong\mathsf{I}$.

To define the isometric or unitary condition of a fermionic tensor $\mathsf{Q}$, our goal is to find another new fermionic tensor $\mathsf{Q}^{\ddagger}$ satisfying $\text{tr}\left(\mathsf{Q}^{\ddagger}\mathsf{Q}\right)=\mathsf{I}$ and $\text{tr}\left(\mathsf{Q}\mathsf{Q}^{\ddagger}\right)=\bar{\mathsf{I}}$ or vice versa depending on the dual of the uncontracted free bond of $\mathsf{Q}$.
$\mathsf{Q}^{\ddagger}$ is the adjoint of $\mathsf{Q}$ and possesses inverted dual for each bond by its nature.
Here trace "$\text{tr}$" means contracting all bonds between $\mathsf{Q}$ and $\mathsf{Q}^{\ddagger}$ except a pair left.
For example, suppose we have a fermionic tensor $\mathsf{Q}$ from QR factorization:
\begin{equation}
    \mathsf{Q}
    =Q_{i_{0}i_{1}i_{2}\cdots}^{m_{0}m_{1}m_{2}\cdots}\vert\mathbf{e}_{i_{0}}^{m_{0}}\rangle\langle\mathbf{e}_{i_{1}}^{m_{1}}\vert\vert\mathbf{e}_{i_{2}}^{m_{2}}.\rangle\cdots
\end{equation}
For example, suppose its $J$-th bond is newly born from the factorization and should be left uncontracted in the isometric trace.
If we define the adjoint
\begin{equation}
    \mathsf{Q}^{\dagger}
    \equiv\bar{Q}_{\cdots{i}_{2}i_{1}i_{0}}^{\cdots{m}_{2}m_{1}m_{0}}\cdots\langle\mathbf{e}_{i_{2}}^{m_{2}}\vert\vert\mathbf{e}_{i_{1}}^{m_{1}}\rangle\langle\mathbf{e}_{i_{0}}^{m_{0}}\vert
\end{equation}
naively, we can compute the left isometry:
\begin{widetext}
\begin{equation}
    \text{tr}\left(\mathsf{Q}^{\dagger}\mathsf{Q}\right)
    =\sum\bar{Q}_{\cdots{i}_{2}i_{1}i_{0}}^{\cdots{m}_{2}m_{1}m_{0}}Q_{i_{0}i_{1}i_{2}\cdots}^{m_{0}m_{1}m_{2}\cdots}\cdots\vert\mathbf{e}_{i_{1}}^{m_{1}}\rangle\langle\mathbf{e}_{i_{0}}^{m_{0}}\vert\vert\mathbf{e}_{i_{0}}^{m_{0}}\rangle\langle\mathbf{e}_{i_{1}}^{m_{1}}\vert\cdots.
\end{equation}
\end{widetext}
Similar for the right isometry $\text{tr}\left(\mathsf{Q}\mathsf{Q}^{\dagger}\right)$.
Obviously, $\text{tr}\left(\mathsf{Q}^{\dagger}\mathsf{Q}\right)\ncong\text{tr}\left(\mathsf{Q}\mathsf{Q}^{\dagger}\right)\ncong\mathsf{I}$ because of extra fermionic signs during the super trace contraction process.
Instead, we choose to define
\begin{equation}
    \mathsf{Q}^{\ddagger}
    \equiv\eta_{J}\cdot\bar{Q}_{\cdots{i}_{2}i_{1}i_{0}}^{\cdots{m}_{2}m_{1}m_{0}}\cdots\langle\mathbf{e}_{i_{2}}^{m_{2}}\vert\vert\mathbf{e}_{i_{1}}^{m_{1}}\rangle\langle\mathbf{e}_{i_{0}}^{m_{0}}\vert
\end{equation}
with attaching an extra sign $\eta_{J}$ as a kind of \emph{graded adjoint operator} to replenish the fermionic supertrace signs~\cite{Scheunert1977}.
It turns out the $\eta_{J}$ also depends on the sides of the conjugation, namely left and right signs:
\begin{equation}
    \begin{aligned}
    \eta_{J}^{[L]}
    &=(-)^{\sum_{j\neq{J}}d_{j}m_{j}}, \\
    \eta_{J}^{[R]}
    &=(-)^{\sum_{j\neq{J}}\bar{d}_{j}m_{j}},       
    \end{aligned}
\end{equation}
where $d_{k}=0, 1$ denotes the type of super vector spaces (normal or dual) of the bond $k$.
$\bar{d}_{k}=(d_{k}+1)\mod{2}$.
Therefore, we say $\mathsf{Q}$ is left- or right-isometric in the sense of this special kind of graded conjugation $Q^{\ddagger}$.

Next let us consider the SVD case, if $\mathsf{M}$ is a fermionic tensor (matrix) and
\begin{equation}
    \mathsf{M}
    =\mathsf{U}\mathsf{S}\mathsf{V}
    \simeq\Tilde{\mathsf{U}}\Tilde{\mathsf{S}}\Tilde{\mathsf{V}}.
\end{equation}
Here $\Tilde{\mathsf{U}}, \Tilde{\mathsf{S}}, \Tilde{\mathsf{V}}$ are truncated according to the singular value spectrum.
If the SVD is not truncated, $\mathsf{U}$ and $\mathsf{V}$ are unitary, which means they are isometric along both row- and column-dimensions, satisfying $\mathsf{U}^{\ddagger}\mathsf{U}\cong\mathsf{U}\mathsf{U}^{\ddagger}\cong\mathsf{V}^{\ddagger}\mathsf{V}\cong\mathsf{V}\mathsf{V}^{\ddagger}\cong\mathsf{I}$.
If the SVD is truncated, $\Tilde{\mathsf{U}}$ and $\Tilde{\mathsf{V}}$ are only isometric along the untruncated dimensions.
That is, only left-graded conjugation of $\Tilde{\mathsf{U}}$ and right-graded conjugation of $\Tilde{\mathsf{V}}$ still preserve the isometric condition: $\Tilde{\mathsf{U}}^{\ddagger}\Tilde{\mathsf{U}}=\Tilde{\mathsf{V}}\Tilde{\mathsf{V}}^{\ddagger}\cong\mathsf{I}$.
However, $\Tilde{\mathsf{U}}\Tilde{\mathsf{U}}^{\ddagger}\ncong\mathsf{I}, \Tilde{\mathsf{V}}^{\ddagger}\Tilde{\mathsf{V}}\ncong\mathsf{I}$.

\begin{figure}
    \centering
    \includegraphics[width=0.3\textwidth]{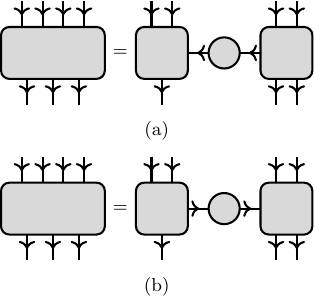}
    \caption{
        \label{fig:ftensor_svd}
        Two different fermionic SVD decompositions of a rank-$\left(3, 4\right)$ fermionic tensor.
        }
\end{figure}

Note that the SVD of a fermionic tensor can be done in another way depending on the arrow direction of $\mathsf{S}$ as shown in Fig.~\ref{fig:ftensor_svd}.
If this case happens, extra signs should be assigned to the isometric fermionic tensors $\mathsf{U}$ and $\mathsf{V}$, respectively.

\section{Some technical comments on the numerical simulations of the $\sigma{t}$-$J$ model}
\label{app:tech-details}

\begin{figure}[t]
    \centering
    \includegraphics[width=0.25\textwidth]{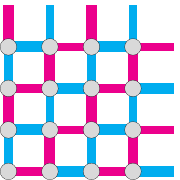}
    \caption{
    Superficial superconductivity patterns with finite $D=12$ from the fTPS simulations of the $\sigma{t}$-$J$ model at $\delta\approx 15\%$.
    $t/J=3.0$.
    Magenta (or cyan) bonds represent the positive (or negative) $\Delta$.
    Bond width is proportional to $|\Delta_{ij}|$.
    }
    \label{fig:sigmatJ_patterns}
\end{figure}

Firstly, as a matter of fact, we find that our fermionic TN method is harder to converge for the $\sigma{t}$-$J$ model in comparison with the $t$-$J$ model.
TN is a factorization for many-body wavefunctions in the real space, which should exhibit area-law of entanglement~\cite{RevModPhys.93.045003}.
The ability of TN strongly depends on the real-space entanglement structure.
For quantum many-body systems with a large Fermi surface, the real-space entanglement is large and harder to be fully described by a TN representation with finite $D$.
This experience agrees well with our results in the main text saying that $\sigma{t}$-$J$ model behaves mostly like a Fermi liquid with a large Fermi surface while $t$-$J$ model is  a non-Fermi liquid.
Finite fTPS probably can provide better convergence, which is left in the future.

Secondly, as we mentioned in our previous work of the $t$-$J$ model~\cite{2411.19218}, we focused on states with uniform charge density and (staggered) magnetization, which is achieved by artificially averaging the bond weights $\Lambda$ during the imaginary time evolution process.
However, for the $\sigma{t}$-$J$ model, we find that taking average on the bond weights $\Lambda$ during the imaginary time evolution process is not the optimal strategy for better energy convergence.
The absence of averaging weights is the numerical reason why we obtain larger error-bars for the superconductivity pairings in the $\sigma{t}$-$J$ model in Fig.~\ref{fig:tJ_sigmatJ_sc}.
By the way, for a finite $D=12$, we could still observe superficial superconducting pairing patterns as illustrated in Fig.~\ref{fig:sigmatJ_patterns}, which looks like the diagonal PDW pattern in the $t$-$J$ model~\cite{2411.19218}.
For $D\rightarrow\infty$, these tiny pairing amplitudes do vanish as we could see in Fig.~\ref{fig:tJ_sigmatJ_sc}.
The power-law behavior of single particle's correlation functions, which is shown in Fig.~\ref{fig:tJ_sigmatJ_cor_up}, also indicates that it is a gapless Fermi liquid without pairing.

\bibliographystyle{apsrev4-2} 
\bibliography{refs}

\end{document}